\documentstyle[12pt]{article}

\input epsf.sty

\def\be{\begin{equation}}
\def\ee{\end{equation}}
\def\bea{\begin{eqnarray}}
\def\eea{\end{eqnarray}}

\newcommand{\eqn}[1]{(\ref{#1})}
\newcommand{\ft}[2]{{\textstyle\frac{#1}{#2}}}

\def\th{{\rm tanh}\Big[\frac{m}{2}(\tau-X)\Big]}
\def\tht{{\rm tanh}^2\Big[\frac{m}{2}(\tau-X)\Big]}
\def\cht{{\rm cosh}^2\Big[\frac{m}{2}(\tau-X)\Big]}

\def\chs{{\rm cosh}^6\Big[\frac{m}{2}(\tau-X)\Big]}
\def\che{{\rm cosh}^8\Big[\frac{m}{2}(\tau-X)\Big]}

\font\cmss=cmss12 
\def\1{\hbox{{1}\kern-.25em\hbox{l}}}
\def\bfZ{\relax{\hbox{\cmss Z\kern-.4em Z}}}

\begin{document}
\begin{titlepage}
\begin{center}
\hfill YITP-SB-00-61 \\
\vskip 30mm

{\Large\bf New instantons in the double-well potential\footnote{Work supported
by NSF grant PHY-972 2101.}}
\vskip 10mm
{\bf Stefan Vandoren and Peter van Nieuwenhuizen}
\vskip 8mm

\centerline{\it C.N.\ Yang Institute for Theoretical Physics}
\centerline{\it State University of New York at Stony Brook}
\centerline{\it NY 11794-3840, Stony Brook, USA}

\vskip 5mm

{\tt vandoren,vannieu@insti.physics.sunysb.edu}

\vskip 8mm

\end{center}

\vskip .2in

\begin{center}
{\bf Abstract }
\end{center}
\begin{quotation}\noindent
A new solution of the Euclidean equations of motion is found 
for the quantum-mechanical double-well
potential with a four-fermion term. It extends the usual 
kink-instanton solution in which both the kink field and the fermionic
fields contain a finite number of new terms which are polynomial 
in the fermionic collective coordinates.
The solution has finite action,
$S=-\frac{m^3}{3\lambda}+\frac{9}
 {140}\,m\,g\,\epsilon^{ijkl}\xi_i\xi_j\xi_k\xi_l$,
where $\xi_i, i=1,...,4$, 
are fermionic collective 
coordinates and $g,m$ and $\lambda$ are coupling constants. We explain why 
in general the instanton 
action can depend on collective coordinates.

\end{quotation}

\end{titlepage}

\section{Introduction}
Instantons are finite action solutions to the Euclidean equations 
of motion in a given field
theory, and describe tunnelling processes between different vacua. 
When the model possesses rigid symmetries, one can act with the 
transformation rules on a solution to generate new solutions, and so
the instanton depends on a number of collective coordinates.
The instanton action is typically inversely proportional to the coupling
constant and does not depend on the collective coordinates generated 
by the rigid symmetries.
Examples are the instanton in pure Yang-Mills theories \cite{Belavin}, or
the instanton in the quantum-mechanical double-well
potential (see e.g. \cite{Rev}).

In supersymmetric theories, or more generally in a theory containing
fermions and bosons, the equations of motion become more complicated
because the fermions can couple non-trivially to the bosonic fields.
An obvious solution is to set all fermions
to zero, and to take for the bosonic fields the usual instanton. 
When rigid supersymmetry is present, one can use the supersymmetry 
transformation rules to obtain other exact solutions of the equations 
of motion with the fermions non-zero and containing fermionic 
collective coordinates.
Because this configuration is related to the purely bosonic one by 
supersymmetry, the instanton action is still the same and does not
depend on the fermionic collective coordinates. 

However, the most general solution to the equations of motion can often 
not be obtained by acting on the purely bosonic configuration with 
symmetries.
As an example consider a chiral spinor, say in the adjoint representation of 
an $SU(N)$ 
gauge theory with $N\geq 3$, satisfying the Dirac equation in the presence 
of a single
instanton. The number of solutions is given by $2N$, so there are
$2N$ Grassmannian collective coordinates (GCC). Such a spinor field is
present in ${\cal N}=1$ $SU(N)$ supersymmetric Yang-Mills theory.
Using ordinary and conformal supersymmetry, one can generate four
GCC, but the remaining solutions of the Dirac equation are not directly 
obtained by symmetry\footnote{However, these other fermionic solutions
 have also been 
given a geometrical interpretation, related to the orientation of the $SU(2)$
subgroup in $SU(N)$ \cite{NSVZ}.}.
Nevertheless, because the 
fermions are just minimally coupled to the gauge fields, the instanton action
is still the same as without the fermions (and hence independent of the GCC), 
because the fermion kinetic term vanishes upon using the Dirac equation.

The question we want to address in this paper is whether the instanton 
action can depend on collective coordinates which are not produced by
any rigid symmetry. Contrary to what is commonly believed, we will show in an 
explicit example that the instanton action can indeed depend on the collective
coordinates of the exact instanton solution. In the discussion we shall comment
on whether this solution can be interpreted as an instanton, and whether 
it is related to other work in the literature.
If we denote a collective
coordinate by $\gamma$ and the fields by
$\phi$, the dependence of the action on $\gamma$ can be computed
from
\be
\frac{\partial S}{\partial \gamma}=\int\, {\rm d}x\, \frac{\partial S}
{\partial \phi(x)}|_{\phi_{{\rm cl}}}\frac{\partial \phi_{{\rm cl}}}
{\partial \gamma}\ ,\label{S-gamma}
\ee
with $\phi_{{\rm cl}}$ the exact instanton solution.
It is then sometimes argued that because
 $\partial S/\partial \phi=0$ is the field
equation, which is satisfied by $\phi_{{\rm cl}}$, 
the action cannot depend on $\gamma$. A closer inspection of \eqn{S-gamma}
however shows that this argument is not correct. The reason is that 
to obtain the field equation, one has to 
do a partial integration. The flaw in the argument is then that one has 
neglected the surface term that appears when doing the partial 
integration: in the instanton background it can yield a non-zero and 
finite contribution. We 
will show this explicitly in a quantum-mechanical model.

In theories where the fermions are not minimally coupled, like the model
we will consider in this paper, or when 
other fields are present which couple non-trivially, the equations
of motion are much harder to solve. Another interesting example is 
the ${\cal N}=4$ supersymmetric Yang-Mills theory, which contains, besides 
the gauge fields,
four Weyl spinors and six scalar fields, all in the adjoint representation.
The equation of motion for the fermions is not the Dirac equation, but
contains inhomogeneous terms due to the presence of Yukawa interactions.
Again, the instanton solution of \cite{Belavin} with all scalars and fermions
vanishing is an exact solution of this model, but one can also include the
fermionic zero modes and try to obtain another exact solution.
Even when the scalar fields have no vacuum expectation value\footnote{If there
is a vacuum expectation value for the scalars, one can use Derrick's theorem
to prove there are no exact solutions to the equations of motion \cite{Der64}.
In such cases, one works with approximate solutions, called constrained
instantons \cite{Affleck}. The instanton action then depends on the collective
coordinates because the field equations are not satisfied.}, it 
is difficult 
to solve the equations of motion with all fermionic collective coordinates
present. 
The strategy one follows is to solve the equations of motion iteratively
in the number of GCC. Because of the Grassmannian nature, this is a finite 
procedure.
For ${\cal N}=4$ super Yang-Mills theory, this procedure was initiated 
in \cite{DHKMV}, and continued in more detail in \cite{BVV}, to lowest order
in the GCC. It was also shown that after one iteration the instanton 
action contains a term proportional to four GCC, see also \cite{DKM}. 
But because of the complicated structure of 
the model (the number of iterations grows with the $N$ of $SU(N)$), an 
exact solution was not obtained. It is important to find the 
complete GCC dependence
of the instanton action, since these terms are needed for computing 
correlation functions beyond the leading order in the  
coupling constant.
Specifically, we have in mind the instanton calculations which are used
to check the AdS/CFT correspondence in ${\cal N}=4$ super Yang-Mills
theory. It is unknown whether a complete solution exists 
in this model, whether it has finite action, and whether the 
winding number is modified.

The purpose of this letter is to use a toy model where this iteration
procedure can be demonstrated explicitly. It is a deformation of the
instanton in a supersymmetric version of the quantum-mechanical 
model with a double-well potential. As we will 
see, the iteration will only take two steps, and hence our construction
leads to a new instanton solution with non-vanishing fermions. We compute
the instanton action, show that it is finite and contains terms proportional 
to four fermionic collective coordinates.

\section{The model}
\setcounter{equation}{0}
We start by considering the following quantum mechanical model,
formulated in Euclidean space,
\be
S_{\rm bos}=-\ft12\int\, {\rm d}\tau \, \{{\dot x}(\tau)^2 + U^2(x)\}\ ,
\ee
where the dot stands for the $\tau$-derivative, and the integration is 
over the infinite $\tau$-line. The potential has a double-well shape,
\be
\ft12 U^2=\ft14 \lambda \Big( x^2 -\frac{\mu^2}{\lambda}\Big)^2\ .
\ee
The field equation, ${\ddot x}-UU'=0$, where the prime stands
for the $x$ derivative, has the usual kink-instanton solution,
\be
x_{K}(\tau)=\frac{\mu}{\sqrt \lambda}{\rm tanh}\Big[\frac{\mu}{\sqrt 2}(\tau-X)
\Big]\ ,
\ee
where $X$ is a collective coordinate associated with time translations. In 
fact, the instanton satisfies
the first-order BPS equation ${\dot x}_K+U(x_K)=0$.

In the spectrum of fluctuations
around the kink-instanton, one finds a zero mode associated with 
translational symmetry,
\be
Z_0(\tau)={\sqrt {\frac{3m}{8}}}\frac{1}{\cht}\ ,\label{bos-zm}
\ee
where we have introduced the variable $m=\mu{\sqrt 2}$. 
This zero mode is obtained by differentiation of the kink-instanton
with respect to the collective coordinate. The normalization is chosen 
such that $\int\,{\rm d}\tau\, Z_0(\tau)Z_0(\tau)=1$.
Evaluating the action on the 
kink-instanton yields
\be
S_{\rm inst}^{(0)}=-\frac{2{\sqrt 2}\mu^3}{3\lambda}=-\frac{m^3}{3\lambda}
\ .\label{S-inst}
\ee
Tunnelling phenomena, including two-loop effects, in this model
were analyzed in \cite{TwoLoop}.

We are interested in models with fermions, so we add the following terms
\be
S_{2-\rm ferm}=-\ft12\int\, {\rm d}\tau \, \{\psi^T_i{\dot \psi}_i
+ (\psi_i^T \sigma_2 \psi_i)U'\}\ .
\ee
Here $i$ is a flavor index, and the fermions are two-component spinors.
For one fermion, there is a supersymmetry, given by
\be
\delta x= \varepsilon^T \sigma^2 \psi\ ; \qquad \delta \psi=\sigma^2{\dot x}
\varepsilon - U \varepsilon\ ,\label{susy}
\ee
but we are interested in the case $i=1,...,4$. The field equations read
\bea
{\ddot x}-UU'&=&\ft12 (\psi_i^T\sigma^2\psi_i)U''\ ,\nonumber\\
{\dot \psi}_i+\sigma^2\psi_iU'&=&0\ .
\eea

The usual instanton solution to this set of equations is obtained by setting
all fermions to zero and taking for $x$ the kink-instanton, $x_K$.
In the supersymmetric case this solution is invariant under SUSY 
transformations with parameter
$\varepsilon=\frac{1-\sigma^2}{2}\varepsilon$, while the parameter
$\varepsilon=\frac{1+\sigma^2}{2}\varepsilon$ breaks SUSY and generates a
fermionic zero mode.

There is another solution, for which the fermions are not zero.
It contains $x=x_K(\tau)$ and the fermion zero modes
\be
\psi^{(1)}_i(\tau)=\xi_i Z_0(\tau) \psi_+\ ,\label{ferm-zm}
\ee
where we have introduced the spinor
\be
\psi_+=\frac{1}{\sqrt 2}\pmatrix{1 \cr i}\ .
\ee
The $\xi_i$ are Grasmannian collective coordinates and $Z_0(\tau)$
is the bosonic zero mode in \eqn{bos-zm}.

The fermionic field equation
is satisfied because $\sigma^2\psi_+=\psi_+$ and ${\dot Z}_0\propto 
{\ddot x}_K$ while ${\dot x}_K+U(x_K)=0$ according to the BPS equation.
 For the bosonic field equation one
notices that the inhomogeneous term on the right hand side 
is zero for two reasons: the product $\xi_i\xi_i$ vanishes but also
$\psi_+^T\sigma^2\psi_+=0$. 
The instanton action for this solution is still given by \eqn{S-inst},
because the fermion terms vanish upon using the field equations.
For $i=1$ the solution in \eqn{ferm-zm} is generated by using the 
broken SUSY in $\delta \psi$ from \eqn{susy}. The normalization is chosen 
such that the zero modes
\be
Z^j_{[i]}\equiv \frac{\partial \psi_i^{(1)}}{\partial \xi_j}\ ,
\ee
are orthonormal,
\be
\langle Z^i| Z^j \rangle \equiv
\sum_k\,\int\,{\rm d}\tau\, (Z^i_{[k]})^*_\alpha (Z^j_{[k]})^\alpha
=\delta^{ij}\ .
\ee

To make the model more interesting, we deform it by adding a
four-fermi term to the action,
\be
S_{{\rm 4-fermi}}=\frac{g}{4}\int\,{\rm d}\tau\, \epsilon^{ijkl}
(\psi^T_i\sigma^1\psi_j)(\psi^T_k\sigma^1\psi_l)\ ,
\ee
where $g$ is a dimensionful coupling constant.
There are actually two terms which are Lorentz invariant in Minkowski 
spacetime: a product of pseudo-scalars (with
a $\sigma^3$ inserted), or an inner product of vector terms, with 
$\sigma^\mu$ inserted. 
One can eliminate the vector terms in favor of the
pseudo-scalar term.

The field equations now take the form
\bea
{\ddot x}-UU'&=&\ft12 (\psi_i^T\sigma^2\psi_i)U''\ ;\qquad
U'=m\,\th\ , \qquad U''={\sqrt {2\lambda}}\nonumber\\
{\dot \psi}_i+\sigma^2\psi_iU'&=&g\,\epsilon_{ijkl}\,\sigma^1\psi_j(\psi^T_k
\sigma^1\psi_l)\ .
\eea
We want to find an exact solution to these equations by iterating in 
the number of GCC. Substituting the fermionic zero mode solution into the right
hand side of the fermionic field equation, one determines the part of 
$\psi_i$ which is cubic in the GCC,
\be
{\dot \psi}_i^{(3)}+\sigma^2\psi_i^{(3)}U'=-g\Big(\frac{3m}{8}\Big)^{3/2}
(\epsilon_{ijkl}\xi_j\xi_k\xi_l)\,\frac{1}
{{\rm cosh}^6\Big[\frac{m}{2}(\tau-X)\Big]}\psi_-\ .
\ee
We used $\psi_+\sigma^1\psi_+=2i$ and introduced the spinor
$\psi_-=-i\sigma^1\psi_+$,
\be
\psi_-=\frac{1}{\sqrt 2}
\pmatrix{1 \cr -i}\ ,
\ee
which is annihilated by $(1+\sigma^2)$. 

Making the ansatz
\be
\psi_i^{(3)}=\alpha(\tau)\, (\epsilon_{ijkl}\xi_j\xi_k\xi_l)\psi_-\ ,
\label{ferm3-zm}
\ee
$\alpha$ is determined by
\be
\dot \alpha-\alpha\,m\, \th =-g\Big(\frac{3m}{8}\Big)^{3/2}
\frac{1}{\chs}\ .
\ee
Since $\cht$ is a solution of the homogeneous equation, we set
\be
\alpha(\tau)=\cht \,y(\tau)\ .\label{alpha-mode}
\ee
The differential equation then reduces to
\be
{\dot y}=-g\Big(\frac{3m}{8}\Big)^{3/2}\frac{1}{\che}\ .
\ee
Changing variables to $z=\th$, we obtain, with $c=-g\,\frac{2}
{m}\Big(\frac{3m}{8}\Big)^{3/2}$,
\be
\frac{{\rm d}y}{{\rm d}z}=c\,(1-z^2)^3\ ,
\ee
which is solved by
\be
y=c\Big(a+z-z^3+\ft35 z^5 -\ft17 z^7\Big)\ ,\label{y-soln}
\ee
where $a$ is an integration constant. 

Looking at the behaviour at $\tau=\pm
\infty$, which corresponds to $z=\pm 1$, we see that $\alpha(\tau=\pm
\infty)$ diverges for generic values of $a$. Even if we choose $a$ such that
$y(z)$ vanishes at $z=-1$ (namely $a=\ft{3}{5}-\ft17=\frac{16}{35}$),
$\alpha(\tau)$ still diverges at $\tau=\infty$, 
\be
\alpha(\tau\rightarrow +\infty)\rightarrow \frac{8c}{35}\,
{\rm e}^{m\tau}\ .
\ee

The solution for $\psi_i^{(3)}$ in the fermionic background 
now acts as a source term for the bosonic field equation. 
Denoting the new term in $x$ proportional to $\epsilon \xi^4\equiv
\epsilon^{ijkl}
\xi_i\xi_j\xi_k\xi_l$ by $x^{(4)}$, one must solve $x^{(4)}$ from the 
equation of motion for $x=x_K+x^{(4)}$,
\be
{\ddot x}^{(4)}-x^{(4)}\,(UU''+U'U')(x_K)=(\psi_i^{(1)T}\sigma^2
\psi_i^{(3)})U''\ .
\ee
Substituting \eqn{ferm-zm}, \eqn{ferm3-zm} and \eqn{alpha-mode} leads to
\be
{\ddot x}^{(4)}-\frac{m^2}{2}x^{(4)}\Big(3\,\tht-1\Big)=-{\sqrt {2\lambda}}
{\sqrt {\frac{3m}{8}}}\,(\epsilon \xi^4)y(\tau)\ ,
\ee
with $y(\tau)$ given in \eqn{y-soln}. 

The homogeneous equation is of course solved by the bosonic zero mode 
$Z_0(\tau)$, so we make the ansatz
\be
x^{(4)}(\tau)=\frac{\beta(\tau)}{\cht}\ .\label{x-four}
\ee
In the same $z$ variable as before, we 
find the following differential equation for $\beta$,
\be
(1-z^2)\frac{{\rm d}^2\beta}{{\rm d}z^2}-6z\frac{{\rm d}\beta}{{\rm d}z}
=\frac{1}{(1-z^2)^2}{\sqrt {2\lambda}}\,(\epsilon \xi^4)g\frac{9}{8m}
\Big(a+z-z^3+\ft35 z^5 -\ft17 z^7\Big)\ .
\ee
The homogeneous equation is solved by
${\rm d}\beta/{\rm d}z=A/(1-z^2)^3$, with $A$ an integration constant, 
which, as we mentioned above, we take proportional to $\epsilon \xi^4$.
So we write $A\,(\epsilon \xi^4)$ instead.
The general solution is then written as
\be
\frac{{\rm d}\beta}{{\rm d}z}=\frac{A\,(\epsilon \xi^4)}
{(1-z^2)^3}+\frac{\gamma(z)}
{(1-z^2)^3}\ .
\ee
This yields a first order differential equation for $\gamma$,
\be
\frac{{\rm d}\gamma}{{\rm d}z}={\sqrt {2\lambda}}\,(\epsilon \xi^4)g
\frac{9}{8m}\Big(a+z-z^3+\ft35 z^5 -\ft17 z^7\Big)\ ,
\ee
which can easily be integrated to give
\be
\frac{{\rm d}\beta}{{\rm d}z}=\frac{(\epsilon \xi^4)}
{(1-z^2)^3}\Big[A+{\sqrt {2\lambda}}\,g\frac{9}{8m}
\Big(az+\ft12 z^2-\ft14 z^4+\ft{1}{10}z^6
-\ft{1}{56}z^8\Big)\Big]\ .\label{eqn-beta}
\ee
One can integrate again,
\bea
\beta(z)&=&(\epsilon \xi^4)\Big(A-{\sqrt {2\lambda}}\,g\,\frac{9}{64m}\Big)
\frac{3}{16}\,{\rm ln}\frac{1+z}{1-z}+\frac{A\,(\epsilon \xi^4)}{8}
\frac{5z-3z^3}{(1-z^2)^2}
\nonumber\\
&&+{\sqrt {2\lambda}}\,(\epsilon \xi^4)\,g\,\frac{9}{8m}
\Big[B+\frac{1}{(1-z^2)^2}\Big(
\frac{a}{4}+\ft{1}{24}(\ft98z+\ft{17}{8}z^3-\ft75 z^5+\ft17z^7)\Big)\Big]\ ,
\label{integrals}
\eea
where $B$ is another integration constant.

It is clear that also $x^{(4)}$ diverges in general at 
$\tau=\pm \infty$ ($z=\pm 1$), but  by choosing $A$ and $a$ appropriately,
$x^{(4)}$ is bounded at infinity, see the discussion in the next section.
 
Notice that we have now 
obtained an exact solution to the full equations of motion. 
Indeed, the iteration in GCC stops at fourth order in the $\xi_i;i=1,...,4$, 
so there is no further contribution to either $\psi_i$ or $x$. The solution is
given by
\be
x=x_K+x^{(4)}\ ,\qquad \psi_i=\psi_i^{(1)}+\psi_i^{(3)}\ ,
\ee
with $x^{(4)}$ given in \eqn{x-four} and \eqn{integrals}, $\psi_i^{(1)}$
given in \eqn{ferm-zm} and $\psi_i^{(3)}$ in \eqn{ferm3-zm}, \eqn{alpha-mode}
and \eqn{y-soln}.

We now compute the instanton action. First of all, the four-fermi 
term gives a contribution
\be
S_{4-{\rm fermi}}=-g\frac{9m}{128}(\epsilon \xi^4)\int_{-1}^{1}\,(1-z^2)^3
{\rm d}z=-g\frac{9m}{140}(\epsilon \xi^4)\ .\label{Sinst-4ferm}
\ee
Next we evaluate the two-fermi terms. Using the field
equation for $\psi^{(1)}_i+\psi_i^{(3)}$, we find that their 
contribution to the action has opposite sign and is twice as large
as the one from the four-fermi term,
\be
S_{2-{\rm fermi}}=g\frac{9m}{70}(\epsilon \xi^4)\ .\label{Sinst-2ferm}
\ee
Finally, we consider the bosonic sector, for which the relevant contribution
to the action can be written as
\be
S_{{\rm bos}}=-\int\,{\rm d}\tau\,\Big[{\dot x}_K{\dot x}^{(4)}
+x^{(4)}U'(x_K)U(x_K)\Big]\ .
\ee
Partially integrating and using the field equations for $x_K$ 
we obtain
\be
S_{{\rm bos}}=-\int\,{\rm d}\tau\,\frac{{\rm d}}{{\rm d}\tau}\Big(
{\dot x}_Kx^{(4)}\Big)=-\frac{m^2}{2{\sqrt {2\lambda}}}
\Big[(1-z^2)^2\beta(z)\Big]^1_{-1}\ ,
\ee
where $\beta(z)$ is as in \eqn{integrals}. This is precisely the surface 
term mentioned in the introduction. So we find
\be
S_{{\rm bos}}=-(\epsilon \xi^4)\Big(\frac{m^2A}{4{\sqrt {2\lambda}}}+
g\frac{9m}{16}\frac{93}{560}\Big)\ .\label{Sinst-bos}
\ee
It is clear that $S_{{\rm bos}}$ depends on the collective coordinate
$A$ but not on $a$ and $B$. The term in $x$ proportional to $A$ is the 
zero-frequency
fluctuation which is not normalizable\footnote{The other zero-frequency
solution is $Z_0$ which is, of course, normalizable.}. It blows up at
$\tau=\pm \infty$, but it contributes a finite amount to the action
because ${\dot x}_K$ vanishes rapidly at $\tau=\pm \infty$.

Collecting the bosonic and fermionic contributions from
\eqn{S-inst}, \eqn{Sinst-bos}, \eqn{Sinst-2ferm} and \eqn{Sinst-4ferm}, we 
obtain the final result for the instanton action
\be
S=-\frac{m^3}{3\lambda}-\Big(\frac{m^2A}{4{\sqrt {2\lambda}}}+
g\frac{261m}{8960}\Big)
\epsilon^{ijkl}\xi_i\xi_j\xi_k\xi_l\ .\label{inst-action}
\ee
Hence although $x$ and $\psi_i$ blow up at $\tau=\pm \infty$, they are regular
at all finite $\tau$ and have finite action.

We are currently studying the relation between this solution and the 
original, purely bosonic one. It may be that one can obtain \eqn{inst-action}
by computing the finite number of tree graphs with four 
fermionic zero modes
on the external legs. The propagator for scalars in the 
background of $x_K$ has been worked out in \cite{TwoLoop} and for fermions
in \cite{PvN}.

\section{Discussion}

We have obtained a new solution of the Euclidean coupled field 
equations of the double
well potential in quantum mechanics with a particular four-fermion 
interaction. The solution depends on four fermionic collective 
coordinates $\xi_i;i=1,...,4$
and contains further three integration constants $A,B,a$, but the action 
is independent of $B$ and $a$. By requiring that the bosonic field $x(\tau)$
remains bounded as $\tau\rightarrow \pm \infty$, we find that 
$a=0$ but $A$ gets fixed to $A=-{\sqrt {2\lambda}}\frac{g}{m}\frac{9\cdot
93}{16\cdot 140}$. For this value of $A$, the contribution \eqn{Sinst-bos} 
from the boson vanishes and the instanton action becomes
\begin{equation}
S=-\frac{m^3}{3\lambda}+\frac{9}{140}\,g\,m\,\epsilon^{ijkl}
\xi_i\xi_j\xi_k\xi_l\ .
\end{equation}

We have called the result of this paper a new instanton,
but is our solution an instanton ? Instantons are 
usually defined as solutions of the Euclidean equations of motion
with finite action. However, one can also define them as configurations 
describing tunnelling processes between different vacua in Minkowski space.
The usual instanton solutions \cite{Belavin,Rev} satisfy both definitions.
Our solution clearly satisfies the first definition. Concerning 
tunnelling, the interpretation of our solution is less clear.
There are vacua with non-vanishing values 
for the fermions, but they are of a different nature from the usual 
vacua. Our solution with $a=0$ and $A$ given above, interpolates between
two configurations with the same energy, but this energy is
non-vanishing and proportional to 
$\epsilon^{ijkl}\xi_i\xi_j\xi_k\xi_l$.
We intend to study the interpretation of our 
solution and whether it describes tunnelling between two such vacua. 

Is our solution new ? We have had detailed discussions with A. Vainshtein 
concerning the relation between our construction 
and the work of V.A. Novikov, M.A. Shifman, A.I. Vainshtein, and V.I. 
Zakharov (NSVZ) on supersymmmetry transformations of collective 
coordinates \cite{NSVZ}. 
Our approach is dynamical; it uses field equations and holds for 
both non-supersymmetric and supersymmetric models. Moreover, we do not 
use spontaneous symmetry breaking and constrained instantons at
intermediate levels. On the other hand, the work of NSVZ starts from
symmetry considerations, in particular supersymmetry and superconformal
supersymmetry. Dynamics 
enters only in sofar as spontaneous symmetry breaking produces a vacuum
expectation value $v$. The supersymmetry transformations act between
both the (bosonic and fermionic) collective coordinates and $v$.
Although constrained instantons are needed in this approach with $v\neq 0$,
one can afterwards take the limit $v\rightarrow 0$ and regain results
which should\footnote{The limit $v\rightarrow 0$ must be taken with 
care, and might not always be the same as starting with $v=0$ \cite{FS}.} 
correspond to the results for unbroken ($v=0$) ${\cal N}=4$ 
supersymmetric gauge theories \cite{DHKMV,BVV}, as was demonstrated 
in \cite{DKM}.
We intend to work out the 
relation between both approaches in the future. In the work 
of NSVZ, supersymmetric gauge theories with $SU(2)$ were considered. The
case of higher $SU(N)$ was treated in \cite{Fuchs}.



\begin{thebibliography}{99}
\bibitem{Belavin}
A. Belavin, A. Polyakov, A. Schwartz and Y. Tyupkin, Phys. Lett. B 59
(1975) 85.\\
G. 't Hooft, Phys. Rev. D 14 (1976) 3432.
\bibitem{Rev}
E. Gildener and A. Patrascioiu, Phys. Rev. D 16 (1977) 423.\\
R. Rajaraman, {\it Solitons and Instantons}, North-Holland,
(Amsterdam, 1982).\\
S. Coleman, {\it The uses of instantons}, in Proc. Int. School of
Subnuclear Physics, Erice (1977), and in {\it Aspects of Symmetry},
Cambridge University Press, (Cambridge, 1985) 265.\\
A. Vainshtein, V. Zakharov, V. Novikov and M. Shifman, {\it ABC of
Instantons}, Sov. Phys. Usp. 25 (1982) 195, and in {\it Instantons in
Gauge Theories}, ed. M. Shifman, World Scientific, (Singapore, 1994).
\bibitem{NSVZ} V. A. Novikov, M. A. Shifman, A. I. Vainshtein and V. I. 
Zakharov, Nucl. Phys. B 229 (1983) 394;
Nucl. Phys. B 229 (1983) 407; Nucl. Phys. B 260 (1985) 157; Phys. Lett. B 217
(1989) 103.
\bibitem{Der64}
G.H. Derrick, J. Math. Phys. 5 (1964) 1252;\\
R. Hobart, Proc. Royal. Soc. London 82 (1963) 201.
\bibitem{Affleck} 
I. Affleck, Nucl. Phys. B 191 (1981) 429. For a recent analysis, see
M. Nielsen and N.K. Nielsen, Phys. Rev. D 61: 105020, 2000, hep-th/9912006.
\bibitem{DHKMV}
N. Dorey, V.V. Khoze, M.P. Mattis and S. Vandoren, Phys. Lett. B 442 (1998)
145, hep-th/9808157.\\
N. Dorey, T. Hollowood, V.V. Khoze, M.P. Mattis 
and S. Vandoren, Nucl. Phys. B 552 (1999) 88, hep-th/9901128.
\bibitem{BVV} A. Belitsky, S. Vandoren and P. van Nieuwenhuizen, Class. 
Quantum Grav. 17 (2000) 3521, hep-th/0004186; Phys. Lett. B 477 (2000) 335,
hep-th/0001010.
\bibitem{DKM} N. Dorey, V.V. Khoze and M.P. Mattis, Phys. Lett. B 396 (1997)
141, hep-th/9612231.
\bibitem{TwoLoop}
A.A. Aleinikov and E.V. Shuryak, Sov. J. Nucl. Phys. 46 (1987) 76;\\
S. Olejnik, Phys. Lett. B 221 (1989) 372;\\
C.F. W\"ohler and E.V. Shuryak Phys. Lett. B 333 (1994) 467, hep-ph/9402287. 
\bibitem{PvN} P. van Nieuwenhuizen, {\it Instantons and tunnelling in
quantum mechanics}, Stony Brook preprint, YITP-SB-00-62.
\bibitem{FS} J. Fuchs and M. G. Schmidt, Z. Phys. C 30 (1986) 161.
\bibitem{Fuchs} J. Fuchs, Nucl. Phys. B 272 (1986) 677; Nucl. Phys. B 282 
(1987) 437.\\
V. A. Novikov, Sov. J. Nucl. Phys. 46 (1987) 554, also in 
{\it Instantons in
Gauge Theories}, ed. M. Shifman, World Scientific, (Singapore, 1994).
\end{thebibliography}
\end{document}